\documentclass[amsmath,amssymb, aip]{revtex4-1}

\usepackage[english]{babel}

\usepackage[letterpaper,top=2cm,bottom=2cm,left=3cm,right=3cm,marginparwidth=1.75cm]{geometry}

\usepackage{amsmath}
\usepackage{graphicx}
\usepackage{xcolor}
\usepackage[colorlinks=true, allcolors=blue]{hyperref}

\begin{document}

\title{Report on the second Toulouse Tensor Workshop}

\author{Jan Brandejs}
\author{Trond Saue}
\affiliation{Laboratoire de Chimie et Physique Quantique, UMR 5626 CNRS — Université de Toulouse, 118 route de Narbonne, F-31062 Toulouse, France }

\author{Andre~Severo~Pereira~Gomes}
\affiliation{Université de Lille, CNRS, UMR 8523 - PhLAM - Physique des Lasers, Atomes et Molécules, F-59000 Lille,  France}

\author{Lucas Visscher}
\affiliation{Department of Chemistry and Pharmaceutical Sciences, Faculty of Science, Vrije Universiteit Amsterdam, De Boelelaan 1108, 1081 HZ Amsterdam, The Netherlands}

\author{Paolo Bientinesi}
\affiliation{Department of Computing Science, Umeå University, Umeå, 901 87, Sweden}

%
%

\begin{abstract}
This report documents the program of the second Toulouse Tensor Workshop which took place at the University of Toulouse on September 17--19, 2025, and summarizes the main points of discussion. This workshop follows the first Workshop (CECAM workshop on Tensor Contraction Library Standardization), which took place in Toulouse one year earlier, on May 24--25, 2024 and led to the formation of a tensor standardization working group, which has since specified a~\href{https://github.com/TAPPorg/tensor-interfaces}{low-level standard interface for tensor operations available freely on GitHub}. The 2025 workshop brought together developers of applications which rely extensively on tensor computations such as quantum many-body simulations in chemistry and physics (material science and electronic structure calculations), as well as developers and experts of tensor software who have the know-how to provide the technical support for such applications. The workshop enabled the community to provide feedback on the specified low-level interface and how it can be further refined. It also initiated a discussion on  how the standardization efforts should be oriented in the near feature, in particular on what should be higher-level interfaces and how to tackle other requirements of the community such as tensor decompositions, symmetric tensors and structured sparsity support.
\end{abstract}

\maketitle

\section{Workshop Background, Aims and scope}


The second Toulouse Tensor Workshop took place from 17 to 19 September 2025 as a follow-up to the first Toulouse Tensor Workshop held in 2024, from which a consensus appeared to emerge that a low-level standard interface for tensor operations could be agreed upon. 

With that, a workgroup (WG) consisting mainly of tensor library developers from academia and industry was established, whose task was to further develop the low-level standard interface and deliver a reference implementation to the community. A report to the workshop participants on the development of this low-level interface, referred to in the following as \textbf{Tensor Algebra Processing Primitives (TAPP)}, is given in~\autoref{tapp-wg-report}. A whitepaper describing \textbf{TAPP} in now available on arXiv~\cite{brandejs2026a}.
 
In view of the advances of the TAPP standardization workgroup (referred to \textbf{TAPP-WG} in the following) over the course of 2024--2025, the main objectives of the 2025 workshop were: 
\begin{itemize}
    \item To gather feedback from the community on the TAPP inteface as well as on the use of its reference implementation;
    \item To initiate a discussion on a higher-level interface than TAPP (which scope such an interface would have, how it could look like etc). Given the scientific profile and corresponding demands of the participants, one particular issue of interest was the handling of sparsity, for example arising from physical symmetries in the application domains;
    \item How concerns about performance (GPU usage etc) should be addressed (if at all) in the standardization effort at this stage.
\end{itemize}
It should be noted that while in the 2024 and 2025 workshops the focus has been on tensor contractions, since these are widespread operations for the applications community, this does not imply other operations to be irrelevant or not amenable to standardization efforts.  Rather, standardization of tensor contractions should be seen as the initial step in such efforts.

\subsection{Workshop organization}

The workshop was structured around three themes, each being the main focus of a day of the meeting: Foundation and concepts (day 1); Libraries, Interfaces and advanced applications (day 2); and HPC, performance portability and future directions (day 3). 

The program for each day consisted of a mix of presentations by participants, panel discussions and breakout sessions. In the breakout sessions participants were split into different groups of about 6-8 members, and brainstormed over questions proposed by the organizers (in part based on the discussions in previous days, or raised in the 2024 workshop and the intervening work by the TAPP-WG), whereas the panel sessions were used to discuss the points of view expressed in the different groups. 

We present the detailed program in appendix~\ref{workshop-program} (including the titles and abstracts for the presentations and the questions by the organizers), and in appendix~\ref{workshop-participants} a list of participants with background information. 

The choice of participants aimed at bringing together representatives of four main stakeholder groups with an interest in tensor operations: (a) low-level (C/C++ based) and high-level (e.g. Python/Julia based) tensor library developers; (b) application experts (scientists developing scientific codes that rely on tensor contractions; (c) hardware vendors that offer mathematical/tensor software for the respective architectures; and (d) high-level mathematical software developers (e.g.\ MATLAB). 


\section{Status reports on 2024 workshop actions}

The two outcomes of the 2024 workshop that were translated into actions were : (1) the establishment of the TAPP-WG to continue work on the definition of a draft low-level implementation aimed at tensor contractions; and (2) the organization of a survey on tensor contractions aimed at establishing a cartography of users and developers of tensor software and their typical use cases.

%
%
%

\subsection{TAPP-WG activities\label{tapp-wg-report}}

The TAPP-WG created in the first 2024 workshop consisted of Jan Brandejs (CNRS, University of Toulouse, France), Devin Matthews (Southern Methodist University, USA), Edward Valeev (Virginia Tech, USA), Paul Springer (NVIDIA), Paolo Bientinesi (Ume\aa~University, Sweden), Justin Turney (Openteams-AI, USA), Alexander Heinecke (Intel, USA) and Christopher Milette (AMD, USA). These were subsequently joined by Lucas Visscher (Vrije Universiteit Amsterdam, The Netherlands), Jeff Hammond (NVIDIA, Finland) and Niklas Hörnblad (Ume\aa~University, Sweden).

As representative of the WG-TAPP, Jan Brandejs reported on their progress in establishing the low-level interface sketched out during the first workshop, and presented what is currently the \href{https://github.com/TAPPorg}{main website for the TAPP-WG}, which contains resources such as \href{https://github.com/TAPPorg/tensor-interfaces}{discussions on the low-level standard tensor interface} and also hosts the \href{https://github.com/TAPPorg/reference-implementation}{Reference TAPP implementation}. 

A whitepaper with a detailed description of TAPP and the reference implementation is now available~\cite{brandejs2026a} and can be consulted for further information.

Apart from the examples provided as part of the hands-on session in how to use the low-level TAPP interface, a first use case of TAPP in a real-life application\cite{fabbroUsingTapp} (involving the definition of Fortran bindings to the C API, and their use on a higher-level interface layer implementing einsum notation) is found in  the~\href{https://gitlab.com/dirac/dirac/-/blob/master/src/exacorr/exacorr_tensor_interface.F90?ref_type=heads}{ExaCorr coupled cluster module}~\cite{Pototschnig2021} of the DIRAC electronic structure code~\cite{saue2020dirac,DIRAC25}.

\subsection{Tensor contraction survey}

The tensor contraction survey was organized by Juraj Hasik (ETH Z\"urich, Switzerland) and Jan Brandejs with assistance from Paolo Bientinesi, with the outcome presented by Juraj Hasik. The data from the survey can be obtained as an appendix to this report or on Zenodo.\cite{surveyShort} The results were presented on the second day of the workshop.

The survey counted 80 respondents out of 130 contacted, the tensor network/condensed matter physics community is predominately represented, followed by the quantum chemistry and quantum computing communities, and with a few respondents from the AI and high-energy physics domains. This is not entirely unexpected since the visibility of this first survey correlates with research fields of the attendants of the 2024 workshop and their professional networks. 

The survey revealed most (69\%) respondents are either developers of application software, users of these, or both. In most cases (69\%) tensor contractions have a strong influence in the performance of the applications, with tensor decompositions being performance critical wholly (18\%) or in part (14\%), and for the majority of respondents tensors with at least three (3) indices typically represent the performance-critical parts of the code. Finally, a minority of respondents store tensors in distributed memory, or use more than a single GPU on a single node (39\% represent single node CPU usage).

For most respondents (67\%), ternary or higher contractions are important use cases, but in most cases such contraction can still be broken down into binary contractions, while the latter are the exclusive use case for 30\% of respondents.

For a high-level interface, Enstein-based notation solutions are the most widely used (43\%), but alternatives are also prevalent. In terms of tensor libraries, Python/Julia based libraries such as NumPy, PyTorch, Itensor, TensorOperations etc are widespread. The survey shows that a majority of respondents (63\%) use automatic differentiation. For most use cases (66\%) a small number of very large tensors is present, but a significant (27\%) number of cases shows a large number ($>1000$) of small tensors. 

With respect to the structure of the data, most use cases either do not possess/exploit sparsity (dense tensors, 34\%), or possess/exploit block sparsity (43\%). The majority of respondents do not employ additional information such as index permutational symmetry, or the presence of thin/skinny tensors, which may have to do with the ability of the underlying tensor software to handle (or not) such information. In line with these findings, a significant portion of respondents do not rely on common sparse formats (48\%).

The results of the survey suggest that at this stage, the community would mostly profit from improvements on handling tensor contractions on single nodes, and improving the ability to exploit block sparsity and the exploitation of index permutational symmetries. They also suggest that one possible way of popularizing TAPP could be by being adopted as the backend of widely used frameworks like NumPy and PyTorch.  One action in this direction has been initiated by Juraj Hasik with the implementation of~\href{https://github.com/jurajHasik/TAPP-torch}{TAPP-torch}, a PyToch operation extension for TAPP.

\section{Presentations' highlights}

The abstracts of the presentations are provided in the appendix, and in the following we present some of the key points evoked by the presenters and from the audience. In order to ensure a broad coverage of subjects, the organizers assigned subjects to all speakers.

\subsection{Pedagogical introductions to tensors and tensor networks}

Since the participants come from different backgrounds and many were not present to the first workshop, the presentations on the first day focused on providing pedagogical introductions to tensor contractions and tensor networks. 

Paolo Bientinesi started out by discussing the efforts to standardize basic linear algebra operations, their success with the definition of the BLAS standard which provided the foundation from which to build more complex standardized software such as LAPACK, and
pointed out the difficulty of higher-level frameworks to achieve efficient use of BLAS/LAPACK. He went on to explain the additional complexities when one should deal with tensor operations instead of matrix operations, and as a result how the current landscape of tensor libraries is extremely fragmented and riddled with reimplementations of the same functionality~\cite{tensorlandscape2022}.
Finally, he outlined prior attempts to standardize tensor operations and how the lessons learned from these motivated the current approach of starting with a much narrower focus on standardizing contractions through an organic, community based effort, and once that is achieved expand the standardization efforts to other operations, such as decompositions.

Lucas Devos provided an complementary view of what tensors are and how tensor networks (a network of connected tensors where each connection represents a tensor contraction) can be represented. Apart from tensor contractions, the presentation highlighted another key operation in tensor networks,  tensor factorizations, which are used to split and compress a tensor network. The presentation also introduced the Penrose diagram notation with which to represent tensor networks, which provides for a simpler representation of contractions or factorizations while exposing the topology of the problem.

\subsection{Library Updates}

For an overview of software for tensor computations, see~\citet{tensorlandscape2022} and references therein.

Devin Matthews outlined the main features of the 2.0 release of~\href{https://github.com/devinamatthews/tblis}{TBLIS} (such as improved complex number support, BLIS plugin to leverage optimized kernels for different CPU architectures, TAPP interface), before explaining how the direct product decomposition (DPD) scheme is used to provide savings in storage and operations by exploiting the block sparsity inherent in the exploitation of point-group spatial symmetry, and how complications arise for higher-order coupled cluster methods, in which tensors may have 6 or more indices are contracted. Mattews also discussed the issue of permutational symmetry (a feature of many tensors appearing in electronic structure) and raised questions as to how/if such symmetries should be handled by TAPP.

Jutho Haegeman presented the suite of tensor libraries (QuantumKitHub) developed in Ghent aiming at providing the building blocks (such as~\href{https://github.com/QuantumKitHub/TensorKit.jl}{TensorKit.jl} and~\href{https://github.com/QuantumKitHub/TensorOperations.jl}{TensorOperations.jl}) for the implementation of tensor network algorithms applicable to many-body physics. Similarly to the case above, a key feature of such systems is the presence of symmetries (Global vs local or internal vs spatial) that introduce block sparsity and index permutational symmetries, and the aforementioned libraries are designed to exploit such symmetries in storage and tensor contractions. In the case of {TensorKit.jl}, dense tensor contractions can be carried out by a TTGT algorithm~\cite{Springer2018:554} and is capable of using different backends such as cuTENSOR, TBLIS or \href{https://github.com/HPAC/TTC}{HPTT}~\cite{Springer2017:910}.

Katharine Hyatt provided an overview of current efforts to support tensor contractions in the Julia ecosystem, which are currently subdivided into QuantumKitHub and ITensor, with emphasis on GPU execution. For NVIDIA architectures, both QuantumKitHub and ITensor use cuTENSOR as backend, and there is currently discussion as to the interest of attempting an implementation of a backend-agnostic Julia "tensor arithmetic" library. Another point discussed is the support of automatic differentiation.

Edward Valeev discussed the~\href{https://github.com/ValeevGroup/tiledarray}{TiledArray framework}, with a focus on how it tackles block sparsity and distributed operations, how tensor operations can be customized at a lower-level as an alternative to a high-level through an eisum-based interface. The presentation also covered a use case for reduced-scaling coupled cluster implementations formulated in terms of nested tensors (tensors of tensors) and how TiledArray handles it.

\subsection{Advanced users and applications experts and high-level interfaces}

Andreas Irmler presented the~\href{https://github.com/cc4s/cc4s}{cc4s} code, which implements coupled cluster theory for solids. A key feature of cc4s is its use of the~\href{github.com/cyclops-community/ctf}{Cyclops tensor framework (CTF)} for carrying out distributed memory calculations.

Kalman Szenes presented the application of DMRG to chemical problems (electronic structure, vibrations etc), and highligted some characteristics of these applications such as the  coexistence of sparse and dense blocks.

Ryan Richard presented the experience of designing~\href{https://nwchemex.github.io/TensorWrapper/}{TensorWrapper}, and the challenges encountered when attempting to support more than a single tensor libraries (currently Eigen is supported, and TiledArray is envisaged).

Jeff Hammond (virtual presentation) discussed his experiences on standardization efforts (such as in MPI), and importance of a layer like TAPP to avoid (suboptimal) re-implementation of infrastructure, such as it occurs in electronic structure theory. He also addressed the challenges of legacy tensor contraction libraries such as the ~\href{https://github.com/sohirata/tce}{Tensor Contraction Engine (TCE)} (and the applications that rely on it) to adapt to changes in paradigm (e.g.\ exploiting GPUs), and how TAPP can/should be integrated into application code (``Design the high-level structure of the application code to map onto TAPP''). He also raised the question of how important tensor operations on distributed memory are nowadays, when state-of-the-art hardware shows capabilities (FLOPs etc) on a single node similar to that of previous generation supercomputers.

Mark Wiebe discussed his experience in~\href{https://github.com/numpy/numpy/commit/a41de3adf9dbbff9d9f2f50fe0ac59d6eabd43cf}{introducing} and developing the \texttt{einsum} package in Python to carry out Einstein summation convention on the operands.

\subsection{HPC performance and portability, HPC experts}

Alexander Heinecke discussed the importance of having reference implementation in order to be able to communicate with Vendors. With respect to performance and in view of the current efforts on the hardware side to support low precision arithmetic as well as toolchains for ``AI'' applications, support for such low precision types becomes relevant, as well as the potential (re)use of these toolchains. At the same time it is important in a standard to limit what is required to be implemented in order to limit the investment by vendors.
 
Alexander Breuer presented work on tensor compilation. 

Grzegorz Kwasniewski provided an overview of~\href{https://developer.nvidia.com/cutensor}{cuTENSOR} and showcased examples of how it is currently being used to handle block sparsity.

Jiajia Li presented an overview of different forms of spartsity in tensors, and how data structures should be adapted to handle these. She also demonstrated how picking a suitable data layout is key to performance, and notably how the use of hash tables for accessing elements can result in significant savings in time.

Christoph Groth introduced the efforts of his team in developing a software stack for tensor contractions and tensor networs in Rust, and discussed the idea of hybrid indexing in tensor networks, where one may exploit partial ordering when this is useful in order to avoid the bookkeeping (in the sequential indexing used by Numpy) or the symbolic indexing (used in \href{link to https://www.itensor.org/}{Itensor}) which may not scale well for large networks.

Örs Legeza presented some highlights of his recent work~\cite{Brower2025} on accelerating large-scale DMRG calculations using mixed precision on Blackwell NVIDIA architectures, where emulation of FP64 via the Ozaki scheme~\cite{10.1007/s11075-011-9478-1} resulted in significant improvements in performance compared to a native FP64 implementation.

Wanqiang Xiong presented an overview of the AMD software stack, and provided a brief description of~\href{https://github.com/ROCm/rocm-libraries/tree/develop/projects/hiptensor}{hipTensor}.

\section{Discussions}


\subsection{Tensor Algebra Processing Primitives}

The first day of the workshop was centered around a presentation of TAPP to the participants, with hands-on exercises as part of the first breakout session. This session helped to detect a few bugs in the reference implementation, as well as to highlight some difficulties with installation and compilation of the reference implementation and of TBLIS, the other backend supported by TAPP (at the time of writing, TAPP also supports the use of cuTENSOR as backend). 

Apart from such issues (which have in the meantime been resolved), the overall assessment of the participants was that as a low-level specification, the TAPP API for tensor contractions is very clear and is on the right track, but (obviously) a higher-level interface is necessary for use by applications developers. The hands-on session also resulted in a request for TAPP was support for carrying out transposition.

Something that was not clear at first for many participants was the nature of TAPP. Instead of recognizing it as the interface to other tensor software, the initial reaction was to view it as another tensor library. Thanks to the hand-on session organized by the TAPP-WG and the discussion after it and in the course of the workshop, this confusion appears to have been completely cleared up.
 
\subsection{High-level interface(s)}

After a first day devoted to the discussion of the low-level TAPP interface, the second day of the workshop focused on the next step(s) to take on the standardization of tensor contractions and how to support them in the standard and in the reference implementation. 

Based on the discussions and the feedback on the presentations, the survey and the participants' needs and interests, the questions of how to add support in TAPP operations on block-sparse tensors residing on a single node appeared as the most useful in the short term, since through it it becomes for instance possible to encode physical symmetries and with that greatly reduce computational cost. 

A preliminary assessment from the participants is that it appears possible to introduce information on block sparsity to the TAPP specification without significant changes in the API, for instance by passing on additional information (such as that currently in use to handle block sparsity by cuTENSOR) to one of be opaque object handles defined by TAPP.

%

Some use cases such as coupled cluster theory in molecular electronic structure may require the distribution of tensors over multiple nodes, depending the size of calculations. At this stage, it was not clear for the participants how to proceed with standardization, or even if standardization is desired or at all possible. The experiences in the development of TensorWrapper as a way to support different tensor libraries as backends to the \href{https://nwchemex.github.io/}{NWChemEx} code serve as a warning with respect to the potential complications. At the same time, a TAPP-based compatibility layer enabling the use of the Cyclops tensor framework for distributed memory tensor contractions has been implemented in the DIRAC code and used in production. This implementation required only very little modification on the side of DIRAC, offering food for thought as how to make TAPP interact with higher-level layers for applications requiring the use of distributed memory.

\section{Conclusions and future directions}

The second workshop on tensor contraction standardization introduced TAPP, an API for low-level tensor contractions on single nodes, and a reference implementation in the C programming language. TAPP currently supports different tensor libraries as backends, and has been showcased in a real-life application, the DIRAC relativistic molecular electronic structure code.

The position of the participants is that a standard interface to tensor contractions is highly valuable to the community, and that TAPP as presented provides a suitable low-level interface to tensor libraries, and can be a pillar over which additional low-level functionality can be built.

The participants considered that incorporating the support for block sparsity into TAPP is the the next step in the standardization of tensor contractions, while it is considered premature to initiate efforts to handle more general forms of sparsity before block sparsity is incorporated. It also seems premature to initiate any standardization efforts for the distributed memory case.  

The opinion of the participants during the workshop is that it should be possible to accommodate such a feature with minimal changes to the current API, and that such efforts should be carried out in a collective manner such as done in the TAPP-WG, either by the current workgroup participants of by the creation of another workgroup with volunteers, and the workshop organizers will help to organize this effort.

Besides the handling of block sparsity, the participants consider that the most suitable interface to be exposed to users is through an einsum-based format, which can be easily mapped to the low-level TAPP as demonstrated for the DIRAC code. 

Finally, many participants share an interest in automatic differentiation and how it can be supported by a standard.  This could be the target of a dedicated workgroup.


\section{Acknowledgments}

This study has been (partially) supported through the EUR grant NanoX n° ANR-17-EURE-0009 in the framework of the "Programme des Investissements d’Avenir". The workshop was partially sponsored by the eSSENCE Programme, under the Swedish Government’s Strategic Research Initiative.

\bibliographystyle{unsrtnat}
\bibliography{workshop}

\clearpage
\appendix
\section{Workshop Program\label{workshop-program}}

\begin{itemize}
\item [\textbf{Day 1}] \textbf{Getting Started \& Foundational Concepts}

\begin{itemize}
    \item[8:30--09:00] Arrival, Nametags \& Introductory Activity 
    
    \item[09:00--09:15] Practical Remarks 
    
    \item[09:15--09:45] Brief summary of the outcome of the preceding workshop [Lucas Visscher and the Working group] 
    
    \item[09:4--10:15] Introduction of Participants 
    
    \item[10:15--10:45] Coffee Break 
    
    \item[10:45--11:15] Pedagogical overview of tensor contractions [Paolo Bientinesi]
    
    \textit{Overview of the field (existing software).}
    
    \item[11:15--12:30] Working group report - Reference implementation, Standardization [Jan Brandejs and the Working group] 

    \item[12:30--13:30] Buffet Lunch
    
    \item[13:30--14:00] Plenary: Instructions for Hands-on Session [Niklas Hörnblad]
    
    \item[14:00--15:30] Hands-on Session in Groups (Coffee available from 14:45)
    
    \item[15:30--16:15] Feedback on the Hands-on session (Panel from the Working group)
    
    \item[16:15--16:45] Pedagogical overview of tensor networks [Lukas Devos]
    
    \item[16:45--18:00] Networking \& Drinks 
    
\end{itemize}

\item[\textbf{Day 2}] \textbf{Libraries, Interfaces \& Advanced Applications}

\begin{itemize}
    \item[9:00--9:15] Practical Remarks

    \item[9:15--9:35]  The Tensor-Based Library Instantiation Software (TBLIS) [Devin Matthews]
 
    \item[9:35--9:55]  Symmetric tensor network algorithms in Julia [Jutho Haegeman] 

    \textit{I will provide an overview of the ecosystem of tensor network Julia packages that we have been developing under the umbrella of the QuantumKitHub organization. This development was started in the Quantum Group at Ghent University, but is gradually turning into an international collaboration. A key feature is the very strong support for symmetries throughout the different algorithms, whereas the more recent focus is on improving support for automatic differentiation and GPUs.}

    \item[9:55--10:15]     Support for tensor operations in Julia libraries [Katharine Hyatt] 

    \textit{The Julia community has developed a variety of libraries for performing low- and high-level tensor operations and implementing algorithms. In this talk I'll summarize the state of Juliaworld for tensor library developers, including addressing automatic differentiation and GPU capabilities.}

    \item[10:15--10:35] TiledArray: generic framework for efficient data-sparse tensor algebra on distributed-memory heterogeneous platforms [Edward Valeev]

    \item[10:15--10:45 ]  Coffee Break

    \item[11:00--11:20 ] High-Performance Tensor Contractions with Cyclops [Andreas Irmler] 

    \textit{This talk explores how large-scale distributed tensor contractions advance quantum chemistry research. After a brief example, it identifies the core challenge: efficiently distributing contractions across thousands of MPI ranks. The Cyclops Tensor Framework is presented as a solution, with a focus on its design, scalability, and performance for these demanding computations.}

    \item[11:20--11:40] The Density Matrix Renormalization Group in Quantum Chemistry [Kalman Szenes] 

    \textit{he density matrix renormalization group (DMRG), relying on the 1-dimensional matrix product state (MPS) tensor network, has established itself as the method of choice for large systems where exact diagonalization procedures are no longer viable. This talk provides an overview of the DMRG algorithm and highlights its applications in quantum chemistry.}

    \item[11:40--12:00] TensorWrapper: Wishful thinking? [Ryan Richard] 

    \textit{The NWChemEx (NWX) team is committed to relying on a series of domain-specific languages (DSLs) to manage the complexity of developing and maintaining a quantum chemistry code. The NWX architecture calls for one of those DSLs to be for expressing tensor operations. Moreover, the design calls for full encapsulation of the performance details. This is a tall order that no current tensor library fully supports. To that end, the NWX team has begun developing TensorWrapper.}

    \item[12:00--12:30] Tensor algorithms and GPU software [Jeff Hammond]

    \item[12:30--13:30] Lunch

    \item[13:30--14:00] Introduction to high-level tensor interface, survey results [Juraj Hasik]

    \item[14:00--15:00] Breakout session--high level interface--structure (Coffee available from 14:30)

    \item[15:00--16:00] Feedback from breakout (panel discussion)--definition of high level interface

    \item[16:00--16:30] NumPy einsum [Mark Wiebe]

    \item[19:15--21:00] Formal Dinner at "Aux pieds sous la table" 
\end{itemize}
\end{itemize}

\begin{itemize}
\item[\textbf{Day 3}] \textbf{HPC, Performance Portability \& Future Directions}
    \begin{itemize}
    \item[09:00--09:40] Pedagogical introduction to performance portability and tensor compilers [Alexander Heinecke]

    \item[09:40--10:00] Tensor Compilation: Exploring Search Spaces [\hyperlink{alex-breuer}{Alex Breuer}]

    \textit{We present the Tiled Execution Intermediate Representation (TEIR), a compact IR for high-performance tensor operations that expresses computation as a composition of primitives over subtensors ("tiles"). TEIR separates what is executed from how it is scheduled through two records: TEIR-Primitives and TEIR-Schedule. The TEIR-Primitives record defines three primitives: first-access, main, and last-access. TEIR-Schedule assigns a per-axis execution policy and specifies axis extents and strides.}

    \item[10:00--10:20] cuTENSOR [Grzegorz Kwasniewski]

    \item[10:20--10:40] High-Performance Sparse Tensor Libraries [Jiajia Li] 

    \textit{This talk will present my group’s research and software infrastructure for sparse tensor operations, covering both element- and block-wise sparsity on CPUs and GPUs.}

    \item[10:40--11:10] Coffee Break

    Presentations: High-Performance Experts

    \item[11:10--11:30] Hybrid indexing of tensor networks [Christoph Groth]

    \item[11:40--12:00] Recent advances in tensor network state methods [Örs Legeza] 

    \textit{A brief overview of recent advances in tensor network state (TNS) methods are presented that have the potential to broaden their scope of application radically for strongly correlated quantum many body systems. Novel mathematical models for hybrid multiNode-multiGPU parallelization on high-performance computing (HPC) infrastructures will be discussed. Scaling analysis on NVIDIA DGX-A100 and DXG-H100 platforms reaching quarter petaflops performance on a single node will be presented. We also report cutting-edge performance results via mixed precision ab initio DMRG Density Matrix Renormalization Group (DMRG) electronic structure calculations, adapted for state-of-the-art NVIDIA Blackwell technology, utilizing the Ozaki scheme for emulating FP64 arithmetic through the use of fixed-point compute resources. Finally, we showcase recent results obtained on IBM superconducting quantum processor with up to 144 qubits, together with classical validation, using state-of-the-art tensor network simulations via a novel Basis Update Galerkin (BUG) method, establishing agreement between quantum and classical approaches. We close our presentation discussing future possibilities via utilization of Blackwell technology in tree-like TNS calculations opening new research directions in material sciences and beyond.}

    \item[12:00--13:00] Lunch

    \item[13:00--14:30] Breakout session: performance portability in context of applications

    \item[14:30--14:50] TBA [Wanqiang Xiong]

    \item[14:50--15:35] Panel discussion on HPC: Anthony Scemama, Justin Turney, Nicolas Renon, Ryan Richard

    \item[15:35--16:00]    Closing Remarks, Feedback and What's next
 \end{itemize}
\end{itemize}

\section{Workshop Participants\label{workshop-participants}}

\begin{itemize}
\item Jan Brandejs, Laboratoire de Chimie et Physique Quantiques, CNRS, Toulouse, France

\textit{Jan Brandejs is a postdoc in the group of Trond Saue specializing on tensor software and code generation for quantum chemistry applications. His main task in the ERC advanced project HAMP-vQED is the creation of programming environment for elegant development of relativistic coupled-cluster methods using distributed memory tensor software for HPC platforms of today.}

\item Paolo Bientinesi, Department of Computing Science, Umeå university, Sweden

\textit{Paolo Bientinesi is professor in High-Performance Computing, and the director of the High-Performance Computing Center North. One of Paolo's research goals is the automatic generation of algorithms and code for linear algebra operations. Together with his research group (HPAC), he has contributed many algorithms and libraries for a range of common tensor operations. \href{https://hpac.cs.umu.se}{hpac.cs.umu.se}}

\item Trond Saue, Laboratoire de Chimie et Physique Quantiques, CNRS, Toulouse, France

\textit{Trond Saue is a CNRS researcher at the Laboratoire de Chimie et Physique Quantiques (Toulouse, France). His current research focuses on the accurate calculation of molecular properties using relativistic coupled cluster theory with the inclusion of QED effects. He is one of the main authors of the Dirac program.}

\item Lucas Visscher, Faculty of Science, Vrije Universiteit Amsterdam, Netherlands

\textit{Lucas Visscher is professor in Quantum Chemistry and Multiscale Modeling at the Vrije Universiteit Amsterdam. He works on developing electronic structure methods with a focus on the description of electronically excited states. His interest in tensor operations arises from his work on coupled cluster theory and quantum computing algorithms.}

\item André Severo Pereira Gomes, Laboratoire de Physique des Lasers, Atomes et Molecules, CNRS, Lille, France

\textit{Andre Gomes is a CNRS researcher at the “Laboratoire de Physique des Lasers, Atomes et Molecules” (Lille, France). His activities focus on the development of electronic structure methods which take into account relativistic effects, electron correlation (primarily with coupled cluster theory) and environment effects (via embedding methods), and their application to simulating molecules containing heavy elements. He is one of the authors of the DIRAC code.}

\item Devin Matthews, Southern Methodist University, USA

\textit{Devin Matthews is an Assistant Professor of Chemistry at Southern Methodist University. His research interests are in electronic structure theory (primarily using coupled cluster methods), computational soft x-ray spectroscopy, tensor factorizations for reduced scaling, and high-performance computing/dense linear algebra. He is one of the authors of the CFOUR program suite.}

\item Edward Valeev, Virginia Tech, USA

\textit{Ed Valeev is Professor of Chemistry at Virginia Tech (Blacksburg, VA, USA). His primary research interests is the development of predictive electronic structure methods with robust control of numerical errors and reduced complexity. Motivated by the needs of such development the Valeev group has been developing a number of software components (Libint(X), TiledArray, MADNESS, SeQuant, BTAS) and as well as leading or contributing to several quantum chemistry packages. Of particular relevance to this workshop are two of such tools: TiledArray, a block-sparse tensor algebra framework for modern distributed accelerated HPC platforms, and SeQuant, a computer algebra system for tensor algebra. In concert they allow rapid composition of conventional and reduced scaling many-body methods.}

\item Jeff Hammond, NVIDIA, Finland

\textit{Jeff Hammond is a Principal Engineer at NVIDIA. He has worked on a wide range of HPC software and hardware projects related to distributed computing, parallel programming models, and numerical algorithms. Jeff received a PhD from the University of Chicago for the implementation and application of coupled-cluster response theory in NWChem.}

\item Jiajia Li, North Carolina State University, USA

\textit{Jiajia Li is an Assistant Professor in the Department of Computer Science at North Carolina State University (NCSU), USA. Her research focuses on high-performance tensor computation, particularly the interaction among applications, numerical methods, data structures, algorithms, and computer architectures.}

\item Grzegorz Kwasniewski, NVIDIA, Switzerland

\textit{Grzegorz Kwasniewski is a Mathematical Libraries Engineer at NVIDIA. He received a PhD at ETH Zurich. His work spans both theoretical and practical aspects of HPC, mostly related to (distributed) linear algebra and data movement optimal algorithms. The applications of interest range from dense linear algebra to "classical" scientific applications and (sparse) distributed LLM training and inference.}

\item Wanqiang Xiong, Advanced Micro Devices, (AMD), Canada

\textit{Wanqiang Xiong is SMTS Software Development Eng. at AMD ML Libraries CAN. He has many years of experiences on HPC software developments, numerical algorithms research etc. Wanqiang Xiong received a PhD from University of Calgary for Chemical and Petroleum Engineering.}

\item Justin Turney, University of Georgia, USA

\textit{Justin Turney is a Senior Research Scientist the University of Georgia in the Center for Computational Quantum Chemistry. His research focuses on theory and code development applicable to combustion related reactions.}

\item Alexander Heinecke, Intel Corporation, USA

\textit{Alexander Heinecke studied Computer Science and Finance and Information Management at Technische Universität München, Germany. In 2010 and 2012, he completed internships in the High Performance and Throughput Computing team at Intel, Munich, Germany and at Intel Labs Santa Clara, CA, USA, working on the Intel MIC architecture. In 2013 he finished his Ph.D. studies at Technische Universität München, Germany. He joined Intel's Parallel Computing Lab in Santa Clara, CA, USA in 2014 as Research Scientist. His core research field is in building a deep knowledge of hardware-aware multi/many-core computing in scientific computing and deep learning. Applications under investigation are complexly structured, normally adaptive, numerical methods which are quite difficult to parallelize. Special focus is hereby given to deep learning primitives such as CNN, RNN/LSTM, Transformers \& MLPs and as well to their use in applications ranging from various ResNets to GPTs in GenAI.}

\item Christopher Millette, Advanced Micro Devices, (AMD), USA

\textit{Chris Millette is a technical lead and development manager at AMD that focuses on implementing GPU libraries for matrix multiplication and tensor primitives. He is a principal contributor to rocWMMA and hipTensor projects for the ROCm software stack, and is highly interested in accelerating applications via instruction and data-level parallelism.}

\item Lukas Devos, Flatiron Institute, USA

\textit{Lukas Devos is a Research Scientist - Software at the Center for Computational Quantum Physics at Flatiron Institute. His research focuses on Tensor Network software libraries based around symmetric tensors, typically using Julia. His current work involves contributing to the ITensors.jl library, and previous work includes TensorOperations.jl, TensorKit.jl, MPSKit.jl and PEPSKit.jl}

\item Ryan Richard, Ames National Laboratory and Iowa State University, USA

\textit{Ryan Richard is a staff scientist at Ames National Laboratory and an adjunct assistant professor of chemistry at Iowa State University. Ryan is the lead developer of NWChemEx, a software environment for developing reusable, modular, and high-performance quantum chemistry tools. A key component of this ecosystem is the TensorWrapper library, which strives to decouple writing a tensor equation from how to performantly solve it. This is done by treating existing high-performance tensor libraries as "physical" layouts and the equation the user writes as a "logical" layout. The guiding design goal of TensorWrapper is for the library to use runtime information about the system and the tensors to choose the ideal mapping from logical layout to physical layout, though admittedly this process is at present entirely manual}

\item Anthony Scemama, Laboratoire de Chimie et Physique Quantiques, CNRS, Toulouse, France

\textit{HPC research engineer. Developer of quantum chemistry programs (Quantum Package, QMC=Chem, ..) and libraries (QMCkl, TREXIO, ...). scemama.githib.io}

\item Jutho Haegeman, Department of Physics and Astronomy, Ghent University, Belgium

\textit{Jutho Haegeman is a professor in quantum many-body physics in the quantum group at Ghent University. Our group develops open-source tensor network packages using the Julia programming language (quantumghent.github.io), with a strong emphasis on exploiting symmetries and studying exotic phases of quantum matter}

\item Sarai Dery Folkestad, Norwegian University of Science and Technology, Norway

\textit{Working as an associate professor at the Norwegian University of Science and Technology. I am interested in development and implementation of electronic structure models and I am a developer of the eT program}

\item Nicolas Renon, Université Toulouse 3 Paul Sabatier, France

\textit{CTO of Toulouse University Computing Center.}

\item Michał Lesiuk, University of Warsaw, Poland

\textit{Michał Lesiuk is a Assistant Professor at Faculty of Chemistry, University of Warsaw, working on development of electronic structure methods based on tensor decomposition techniques (group webpage: aesmgroup.chem.uw.edu.pl).}

\item {Alex Breuer\label{alex-breuer}, Friedrich Schiller University Jena, Germany}

\textit{Alex Breuer leads the Scalable Data- and Compute-intensive Analyses lab. His research and teaching activities cover the full spectrum of algorithms and software for modern and emerging hardware. In 2014, Alex was honored with an ACM/IEEE-CS George Michael Memorial HPC Fellowship for his Ph.D. project “High Performance Earthquake Simulations”. In addition, he and his collaborators have been awarded with the PRACE ISC Award and nominated as ACM Gordon Bell finalists. Alex holds a doctoral degree from the Technical University of Munich.}

\item Örs Legeza, Wigner Research Centre for Physics, Hungary

\textit{Örs Legeza is a professor at Wigner Research Centre for Physics and a Hans Fischer Senior Fellow at the Institute for Advanced Studies at the Technical University of Munich. His research interests focus on the development of novel tensor network state (TNS) methods and their application to strongly correlated quantum many-body systems to simulate and study magnetic properties in solid states, exotic quantum phases, complex molecular clusters, ultracold atomic systems, and nuclear structures. For the method development he combines established methods for simple networks (MPS, MERA, tensor trees) with concepts from quantum information theory and computational mathematics to push the current frontier of moderate system size to much larger and more complex systems via simulations on high performance computing infrastructures.}

\item Katharine Hyatt, University of Ghent, Belgium

\textit{Katharine Hyatt is a postdoc at UGent working on GPU acceleration of the tensor software here as well as AD methods for tensor network algorithms. She is also interested in tensor networks as a tool for simulation of near term quantum hardware and quantum algorithms in general.}

\item Juraj Hasik, University of Zurich, Switzerland

\textit{Juraj Hasik is a postdoc at the University of Zurich, within the Quantum Matter Group led by prof. Neupert. His research centers on the development and application of tensor network methods to study strongly correlated systems in two dimensions. He develops peps-torch and YASTN libraries, which are designed to optimize two-dimensional (symmetric) tensor networks.}

\item Matthieu Mambrini, Laboratoire de Physique Théorique, Toulouse, France

\textit{Matthieu Mambrinbi is a CNRS researcher at the Laboratoire de Physique Théorique (Toulouse, France). His research focuses on quantum magnetism and spin liquids using various numerical methods. He is involved in tensor network methods and more particularly in the PEPS scheme, in a context where continuous (e.g. SU(N)) and discrete (e.g. point group) symmetries play a central role.}

\item Kalman Szenes, ETH, Switzerland

\textit{Kalman Szenes is a PhD student in the theoretical chemistry group of Prof. Markus Reiher at ETH, Zurich. He is one of the developers of the QCMaquis density matrix renormalization group program. His research focuses on method development for electronic structure calculations on strongly correlated molecules. Personal webpage: kszenes.github.io}

\item Andreas Irmler, Institute for Theoretical Physics, TU Wien, Austria

\textit{Andreas Irmler is a postdoctoral researcher at TU in Vienna. As a member of Andreas Gruneis' group, his work focuses on developing new theoretical methods and computational implementations for large-scale coupled-cluster calculations. He is a co-author of the coupled-cluster code CC4S, which enables massively parallel computations across hundreds of compute nodes}

\item Marco Trenti, Tensor AI Solutions GmbH, Germany

\textit{Marco Trenti is the CTO of Tensor AI Solutions, a startup focused on Explainable AI (XAI) and tensor network methods. His main work involves developing high-performance computing (HPC) algorithms for tensor network machine learning (TNML) and researching their applications in data science, quantum computing, quantum simulations, and industrial problems. He is particularly interested in specific contraction patterns and tensor algebra kernels that appear in TNML.}

\item Niklas Hörnblad, Department of Computing Science, Umeå university, Sweden

\textit{Niklas Hörnblad is a project assistant for Paolo Bientinesi at Umeå University. He made his master's thesis about tensor operations and has continued working in that area with Paolo Bientinesi. He is the main contributor to the reference implementation of TAPP.}

\item Yannick Lemke, Ludwig-Maximilians-Universität München, Germany

\textit{Yannick Lemke is a PhD student in the group of Prof. Christian Ochsenfeld at LMU Munich working on method development for electronic structure theory. His research topics include electron correlation methods based on the random phase approximation as well as the development of the constraint-based orbital excited state method. Yannick is one of the authors of the FermiONs++ program package}

\item Christoph Groth, Pheliqs/IRIG/CEA Grenoble, France

\textit{Christoph Groth is a staff researcher at the theory group of the quantum physics research lab "Pheliqs" of CEA Grenoble. His fascination with computing was kindled by a ZX Spectrum at the age of 6 years. Christoph’s efforts to widen his horizon by studying physics could never obscure this first love. Even now, he is more into finding the best ways to represent physical problems inside a computer than solving the problems themselves. Nowadays, Christoph works mostly on tensor network methods (algorithms, APIs, and libraries). He is one of the authors of the "Kwant" library for quantum transport computations}

\item Alejandro Estaña, CNRS - Université Toulouse 3 Paul Sabatier, France

\textit{HPC and AI Research Engineer in the Application Support Team at CALMIP Computing Center.}

\item Johann Pototschnig, CASUS, Helmholtz Zentrum Dresden-Rossendorf, Germany

\textit{Johann Pototschnig is a PostDoc in the Scientific Computing Core group of Andreas Knüpfer at the Center for Advanced Systems Understanding. He is working on integrating new interfaces to the CP2K code on the performance of the code. His research focuses quantum chemical methods for HPC infrastructure. Previously he contributed to the massively parallel coupled cluster in DIRAC.}

\item Laurenz Monzel, Saarland University, Germany

\textit{PhD student in the group of Prof Stella Stopkowicz, Saarland University, Germany. Method developing of QED Coupled Cluster wave functions which treat the electronic part and the EM field part on an equal footing. This allows the accurate description of light-matter hybrid states and requires an efficient algebra and storage of fermionic and bosonic tensors.}

\end{itemize}

\clearpage

\section{Breakout session questions\label{workshop-breakout-quesitons}}

\begin{itemize}
\item 18 September Session
    \begin{itemize}
    \item Quick questions (not covered by survey)
        \begin{itemize}
            \item How much supercomputer node hours do you use annually ?
            \item Which datatypes are crucial for you ?
        \end{itemize}
    \item Evaluation of use of TAPP
        \begin{itemize}
            \item How would you design a wrapper or a high-level interface tailored for your environment?
            \item Would you use code generators in this interfacing ?
        \end{itemize}
    \item What’s next ?
        \begin{itemize}
            \item What are the most important features currently missing in TAPP ?
            \item Should a working group provide a sample high-level wrapper of TAPP on separate repository to set best practices?
            \item Ingredients for a standardized high level interface built on top of TAPP ?
        \end{itemize}
    \end{itemize}
\item 19 September Session
    \begin{itemize}
    \item Seminar room group
        \begin{itemize}
            \item How can TAPP best deal with tensors stored in distributed memory ?
            \item Do we need a standardised interface for tensor initialization ? If yes: what should it look like ?
        \end{itemize}
    \item Other groups 
        \begin{itemize}
            \item How should a function call for tensor transposition look like ?
            \item How do you encode the sparsity information into a tensor contraction call ?
            \item How should a function call to create a block-sparse tensor look like ?
        \end{itemize}
    \item all groups: How should we organize ourselves after this meeting ?
    \end{itemize}
\end{itemize}



\clearpage



\end{document}